\documentstyle[12pt]{article}

\newcommand{\bce}{\begin{center}}
\newcommand{\ece}{\end{center}}
\newcommand{\beq}{\begin{equation}}
\newcommand{\eeq}{\end{equation}}

\newcommand{\Eq}[1]{Eq.~(\ref{eq:#1})}       
\newcommand{\eq}[1]{(\ref{eq:#1})}           

\newcommand{\refem}{\em}  

\newcommand{\PRL}[1]    {{\refem Phys.\ Rev.\ Lett.\/} {\bf #1}}

\newcommand{\PRA}[1]    {{\refem Phys.\ Rev.\/} {\bf A#1}}

\newcommand{\JSP}[1]    {{\refem J.\ Stat.\ Phys.\/} {\bf #1}}

\newcommand{\JPA}[1]    {{\refem J.\ Phys.\/} {\bf A#1}}

\newcommand{\PHYSA}[1]  {{\refem Physica\/} {\bf A#1}}

\newcommand{\NPB}[1]    {{\refem Nucl.\ Phys.\/} {\bf B#1}}

\newcommand{\PMB}[1]    {{\refem Phil.\ Mag.\/} {\bf B#1}}
\newcommand{\CP}[1]     {{\refem Contemp.\ Phys.\/} {\bf #1}}

\newcommand{\NAT}[1]    {{\refem Nature\/} {\bf #1}}

\newcommand{\IJMPC}[1]  {{\refem Int.\ J.\ Mod.\ Phys.\/} {\bf C#1}}

\begin{document}

\title{\vspace{-4.0cm} {{\normalsize HLRZ preprint 26/93} \hfill}
        \vspace{2.0cm} \\
        Statistical Laws and Mechanics of Voronoi Random Lattices}

\author{Kent B\ae{}kgaard Lauritsen,\hspace{-1.5mm}* 
	~Cristian Moukarzel \\
	and Hans J. Herrmann \\
        {\em HLRZ, Forschungszentrum J\"ulich}, \\
        {\em Postfach 1913, D--5170 J\"ulich, Germany}
        }

\date{}

\maketitle

\begin{abstract}
We investigate random lattices where the connectivities are
determined by the Voronoi construction, while the location
of the points are the dynamic degrees of freedom.
The Voronoi random lattices with an associated energy are immersed
in a heat bath and investigated using a Monte Carlo simulation algorithm.
In thermodynamic equilibrium we measure coordination number
distributions and test the Aboav-Weaire and Lewis laws.
\end{abstract}

%
%



\setlength{\baselineskip}{0.7cm}

\newpage

\section{Introduction}

Random lattices are a convenient way of
approximating cellular network structures \cite{weaire-rivier:1984} and of
discretizing space without introducing any kind of anisotropy
\cite{christ-etal:1982}.
The application of random lattices therefore
covers a great variety of physical systems
spanning from membrane and vesicle research, soab bubbles, sphere packings,
kinetic growth models, large-scale structure of the universe
to quantum field theory and two-dimensional
quantum gravity
\cite{kroll-gompper:1992,moukarzel:1993a,oakeshott-edwards:1992,moukarzel:1992,coles:1990,boulatov-etal:1986,benav-etal:1992}.
%
%
It is therefore important to investigate and characterize the
properties of random lattices and cellular structures.
For cellular networks, e.g., grain mosaics, plants and foams
relations such as the Aboav-Weaire and Lewis
laws have, empirically, been found to provide good statistical descriptions.
The Aboav-Weaire law states that the total
number of neighbours of cells neighbouring an $n$ sided cell is linear
in $n$, and has, recently, been derived from maximum entropy arguments
\cite{peshkin-etal:1991}.
However, in the case of a purely topological model it
has been shown that
there are small corrections to the linear relation \cite{godreche-etal:1992}.
The Lewis law, which states that the average area
of $n$ sided cells is linear in $n$, has previously
been shown to follow from maximal arbitrariness in the cellular
distribution \cite{rivier-lissowski:1982}.

In the present paper we investigate two-dimensional random lattices
where the location of the points are the dynamic degrees of freedom and
the connectivities are at all times
determined by the Voronoi construction.
The purpose is twofold: First we introduce random lattices which will
be in thermodynamic equilibrium and, secondly, this type of random lattices
is then investigated in order to test statistical laws.
The Voronoi construction or tessellation
is defined as follows: For all points determine the
associated cell consisting of the region of space nearer to the point
than to any other point.
Whenever two cells
share an edge they are considered as neighbours and one draws a link
between the two points associated to/located in the cells.
In this way one obtains the triangulation of space that is
called the Delaunay lattice. The Delaunay random lattice
is dual to the Voronoi tessellation in the sense that points
corresponds to cells, links to edges and triangles to the vertices
of the Voronoi tessellation.
The triangulation (tessellation) of space with $N_P$ points (cells),
$N_L$ links (edges)
and $N_T$ triangles (vertices) are constrained
by the Euler relation
\beq
	N_P - N_L + N_T = \chi
							    \label{eq:euler}
\eeq
where $\chi$ is the Euler characteristic, which equals 2 for a graph
with the topology of a sphere and 0 for a torus.
Since $3 N_T = 2 N_L$ always holds for a triangulation,
one obtains $N_T = 2N$ and $N_L = 3N$ for $N_P = N$ points
on a torus.
The connections of the lattice determine
the connectivity matrix $C_{ij}$ which is 1 if the points $i$ and $j$ are
connected with a link and 0 otherwise.
Voronoi structures have previously been the subject of interest
in computer-simulation studies in two dimensions
\cite{lecaer-ho:1990,glaser-clark:1990,diane-etal},
where the Voronoi cells were used to determine nearest neighbours,
and three dimensions \cite{kumar-etal:1992}.
Recently, the concept of Voronoi
tessellations has been generalized in order to model
equilibrated soap froths \cite{moukarzel:1993a}.
In the study in Ref.~\cite{diane-etal} the Voronoi tessellation was
dynamically maintained during the simulation, and used to describe
structural changes in the hard-disk model
but the Voronoi construction was not used to determine the dynamics.

In the next section we define the dynamic random lattice.
In Sec.~3 we present simulation results, and in Sec.~4 we
investigate statistical laws. Finally, in section 5, we
summarize and conclude.

\section{Dynamic Random Lattices}

In our work the Voronoi construction will dictate the dynamics since
we are interested in the dynamic properties of random lattices.
To the random lattice we therefore assign a
thermodynamic energy functional of the general form
\beq
	H = \sum_i E[n_i, \sigma_i, \ell_{ij}, C_{ij}]
							\label{eq:energy-func}
\eeq
where $n_i$ is the coordination number of the point $i$,
$\sigma_i$ the area of Voronoi cell number $i$,
and $\ell_{ij}$ the distance between the points $i$ and $j$.
The lattice is taken to be in equilibrium with a heat bath at
temperature $T= \beta^{-1}$ (we use units where Boltzmann's
constant equals unity).
Depending on the actual energy functional~\eq{energy-func}
assigned to the lattice
very differently ordered states may appear at low temperatures.
The form of the energy is not neccessarily
supposed to represent a description of a physical system; of course
one can construct physically motivated actions as well but this
is not our interest in this study.

At high temperatures the random lattice will be of Poissonian type
\cite{christ-etal:1982,lecaer-ho:1990}, i.e.,
the positions of the points are uncorrelated and
uniformly distributed over the entire lattice.
On the other hand,
the action can be ingeniously chosen so that one obtains, at
low temperatures, lattices with certain useful properties.
In, e.g., the numerical growth of Diffusion-Limited Aggregation (DLA)
clusters by solving the Laplace's equation on a random lattice,
\mbox{$-(\Delta \phi)_i = \sum_j C_{ij} \sigma_i (\phi_i - \phi_j)/ \ell_{ij}$}
\cite{moukarzel:1992},
the convergence properties can be speeded up
when the couplings
$\sigma_i / \ell_{ij}$ are nearly constant.
Such a lattice may be obtained
by using an appropriate energy~\eq{energy-func}
and then equilibrating the random lattice in a Monte Carlo simulation.

With the standard Monte Carlo simulation techniques we have
investigated Voronoi random lattices. Corresponding to the usual spin-flip for
an Ising model the points defining the lattice
are subject to random displacements. The new lattice is also determined
by the Voronoi construction
and the move is accepted with the Metropolis probability
\beq
	P = \min \left( 1, \, e^{- \beta \Delta H} \right)
							\label{eq:metropolis}
\eeq
where $\Delta H$ is the increase in energy produced by the move.
Provided the
detailed balance condition is fulfilled the Monte Carlo
dynamics will correctly describe a Voronoi random
lattice in equilibrium with a heat bath at temperature $T$.
By means of simulations
one can measure thermodynamic quantities as well as testing statistical
laws supposed to be valid for an equilibrated graph itself.

In Fig.~1 we show how a typical change of the lattice caused by the
move of one point to a random nearby position looks like.
[Another way to determine the new position is to
restrict the point to, e.g., the Voronoi cell
around the point---or to within a fixed distance, $r$, from the edges
of the Voronoi cell; in this latter way non-overlapping particles of radius
$r$ (e.g., granular material \cite{moukarzel:1993b})
can be simulated with a hard-disk
interaction (similar to the model studied in Ref.~\cite{diane-etal}).]
When the point $P$ is moved
as shown in Fig.~1 a link at the left is
``flipped''---a neighbour-loosing process---and a link at the right
is flipped---a neighbour-gaining process---in order that the
Voronoi construction is fulfilled for the random lattice
with the point in the new position.
No matter how small the move of the point is taken to be, very
complicated changes which can influence a large
part of the lattice can take place.
Even though it is impossible in advance to predict how profound
the changes will be, it is always
possible to view such complicated changes as a combination
of neighbour-gaining and loosing processes
and in this way keep track of how the lattice gradually changes
as a point is moved.
When the new position is very close to the old position, simple
changes, as shown in Fig.~1, are, however, most likely to take place.

We have studied a variety of different Hamiltonians, typically of the form
\linebreak 
\hbox{$H = J \sum_i F_{i}^{2}$}, with $\sum_i F_i = const.$
(e.g., $F_i = n_i, \sigma_i, \ldots $), where the sum is over the $N$ points.
A Hamiltonian of this kind with positive coupling constant $J$
(which for matter of convenience is chosen to be unity) will favour
a low-temperature phase where the quantities $F_i$ are all equal
due to the imposed constraint $\sum_i F_i = const$.
The case of negative $J$, on the other hand,
will have low-temperature states where the $F_i$'s are small at some points
and large at others.
Here we focus on the case where the energy depends
on topological quantities and study the Hamiltonian
\beq
	H = J \sum_{i=1}^{N} (n_i - 6)^2
							      \label{eq:H=q2}
\eeq
with positive and negative coupling constants;
elsewhere results on other models,
including couplings to Ising spins, will be given.
One notices that,
since a torus imposes the constraint $\sum_i n_i = 6 N$ on the
coordination numbers (from \Eq{euler}), the same
dynamics is obtained with, e.g., $H = J \sum_i n_{i}^{2}$; in both cases
the Hamiltonian~\eq{H=q2} will favour that all $n_i$ equal 6.

\section{Simulations}

In the Monte Carlo simulations we use a lattice with the topology of a torus,
i.e., we impose periodic boundary conditions in both directions.
The lattices we have simulated contained from $N = 504$ to 1254 points.
One Monte Carlo step (MCS) corresponds to one attempted move
per particle. For each temperature we carried out 1000--2000 MCS
to reach equilibrium and then measured data for another 10,000 MCS.
Since we are interested in the properties of the random
lattices themselves we use point updates where the new position
of a point is chosen within a box surrounding the old position.
The size of this region is chosen
so that the acceptance rate is about 50\% in an intermediate temperature
interval and not too small for low temperatures.
As previously remarked, it is not possible in advance to know
the amount of change that will take place when a point is moved.
This implies that the updating algorithm is intrinsically
non-parallelizable/non-vectorizable. Therefore, workstations
(Sun Sparc 10's)
have been used in our simulations of the dynamic
random lattices. We have implemented different versions of the algorithm
using both index tables in Fortran and dynamic pointer structures in
the C programming language (see, e.g., Ref.\ \cite{kbl-etal:1993}). Typically,
we obtain an updating time for one point-moves of the order of 1 msec
(the time scales approximately linearly
with the size of the box in which the points are allowed to be moved).

The positive coupling Hamiltonian~\eq{H=q2} has as
groundstate a lattice with the topology of a triangular lattice.
In the phase space of positions of points one has in fact many
groundstates, namely all distorted triangular lattices,
since for the regular triangular lattice a point can be moved to a
certain amount without leading to a link-flip.
These short displacements of the points are not
detected by the Hamiltonian~\eq{H=q2} which measures only the
coordinations of the random lattice.
A low-temperature lattice ($T = 0.1$)
is shown in Fig.~2(a), and in Fig.~2(b) is shown
the (dual) Voronoi tessellation for the same configuration of the points.
One notices that the points do not seem to order
in the low-temperature phase. Nevertheless, 96\%
of the points have
$n_i = 6$ for the lattice shown in Figs.~2(a)-(b).
Measurements of the energy as a function of temperature show
that the energy is a steadily increasing function and the specific heat
$c(T) = V^{-2} T^{-2} \left( \langle E^2 \rangle - \langle E \rangle^2 \right)$
determined by means of the fluctuation-dissipation theorem as the
fluctuations in energy shows no sign of a peak except at zero temperature
with trivial exponent. These findings are in agreement with the
Mermin-Wagner theorem \cite{mermin-wagner:1966} which states that
a two-dimensional continuous system cannot undergo an
order-disorder phase transition at a finite temperature.

Starting from an almost regular triangular lattice at a low temperature
and heating up the system one observes
single localized link flippings which create points with 5 and 7
neighbours respectively, i.e., pairs of ``defects''.
Such points typically merge after a few Monte Carlo steps
but as the temperature is increased these defects start to break up
and the lattice becomes more and more random until it, eventually, will
end up as the Poissonian random lattice at infinite temperature.
On the contrary,
cooling the high-temperature lattice these defects
merge (annihilate) and eventually the points with 6 neighbours
``percolate'' from one side of the system to the other.
Due to the periodic boundary conditions some topological defects
can be ``frozen in''
in the low-temperature ``ordered'' phase where, e.g., connections
across a line have all been shifted by one or more units.
Qualitatively this scenario ressembles the unbinding of vortices
typical of the Kosterlitz-Thouless transition of two-dimensional melting.
One way of describing these
low-temperature patterns is by identifying
the characteristic ``ABC-sublattice'' ordering for a triangular lattice
and to see if this order propagates all over the system
(i.e., percolates);
results of such an analysis will be described elsewhere.

The negative $J$ Hamiltonian~\eq{H=q2} will have as its groundstate
a lattice where one point has all the other points as neighbours,
since the energy~\eq{H=q2}
favours an uneven distribution of the number of neighbours.
For low temperatures there will be many metastable disordered states
one of which is shown in Figs.~2(c)-(d).
Accessing these states by simulated annealing we find that
up to 80--90\% of the points have $n_i = 4$ and these points
lie along (fluctuating) lines.
Specifically, the lines tend to close in circular patterns with a
many-neighbour point
(e.g., $n_i \, \,^{>}_{\sim} \, 60$)
lying in the middle, cf.\ Figs.~2(c)-(d).
{}From Fig.~2(c) one notices that the circular patterns are roughly of
the same size.
Due to the very low energy associated to these central points
it is highly unlikely that
once a circular pattern has emerged it will break up again.
The system therefore freezes into a metastable state and
high energy barriers must be overcome to get out
(at high temperatures these patterns will not appear but for low temperatures
one has to be aware of the possibility of ending up in
a metastable state and not the true equilibrium state).
We have measured the energy as a function of time. For high temperatures
there is a fast relaxation to equilibrium but as the temperature is
lowered to $T \approx 11$ the system enters a new phase where the relaxation to
equilibrium is very slow (possibly determined by a power-law behaviour)
being due to the many low-temperature disordered metastable
states in the system.
In fact it is possible that in the low temperature regime the system
always evolves towards the topologically unique groundstate
given by all $n_i = 4$ except one which has $n = 2N$ but that
the systems evolution becomes extremely slow like in a glass.

\section{Statistical Laws}

{}From the Euler relation~\eq{euler} follows that the average
coordination number of a graph on a torus is 6, i.e.,
\beq
	\langle n \rangle = \sum_n n P_n = 6
							\label{eq:n=6}
\eeq
where $P_n$ is the (normalized) probability, $\sum_n P_n = 1$,
of finding a point with $n$ neighbours
(this result is true for a graph of arbitrary topology in
the limit $N \to \infty$, since the Euler characteristic $\chi$
is a number of order unity).
This probability density has been plotted
for various temperatures in Fig.~3 for the two models studied.
At low temperatures
for the positive coupling model one notices a strongly peaked density at
the value $n=6$ reminiscent of the topology of the triangular lattice.
Remember, however, that the lattice does not have the regular structure
of a triangular lattice and even at rather low temperatures already
contains defects and irregularities (cf.\ Figs.~2(a)-(b)).
The variance (the energy per point)
\mbox{$\mu_2 = \left< (n-6)^2 \right> = \sum_n (n-6)^2 P_n$}
decreases as a function of temperature and
approaches zero at low temperatures, where only points with 5, 6 and 7
neighbours are present.

In the negative coupling constant case Fig.~3(b) shows that
owing to the points with many neighbours (cf.\ Figs.~2(c)-(d))
the $P_n$ density is broader for low temperatures as compared to the
case of positive $J$.
As the temperature is lowered the variance $\mu_2$
increases from the Possonian random lattice value \mbox{$\mu_2 \approx 1.78$}
(see, e.g., Ref.~\cite{lecaer-ho:1990})
to \mbox{$\mu_2 \approx 2.7$} at the
temperature $T = 11$, where the circular patterns start
to form. The occurence of the very high $n$ values makes
$\mu_2$ increase even further as the temperature is decreased.
{}From Fig.~2(b) one notices the change in the overall form of $P_n$
taking place at the transition temperature $T \approx 11$.
The transition seems very sharp and it would be interesting to
investigate if it is of first or second order.

We now turn our attention to the correlations present
in the equilibrated random lattices.
Denoting by $m(n)$ the average number of neighbours of the neighbours
of a cell with $n$ neighbours
the Aboav-Weaire law states that (see, e.g, Ref.~\cite{weaire-rivier:1984})
\beq
	n m(n) = (6-a) n + (6a + \mu_2)
							\label{eq:aboav-weaire}
\eeq
with $a \sim 1$, empirically.
Note that by parametrizing the Aboav-Weaire law by $a$ as done in
\Eq{aboav-weaire} the sum rule
$\langle n m(n) \rangle = \langle n^2 \rangle = \mu_2 + 36$
is automatically fulfilled \cite{weaire-rivier:1984}.
This law has recently been predicted from
maximum entropy arguments \cite{peshkin-etal:1991},
whereas the analytical solution of a topological model showed
that the linear law is only approximately true
\cite{godreche-etal:1992}.
For the Poissonian random lattice it is well established
that the linear Aboav-Weaire law does only approximately
hold \cite{lecaer-ho:1990,rivier:1985}.
This has been explained by arguing that this
particular random lattice is not fully equilibrated---it is young
in the terminology of statistical crystallography and is
space-filling by construction rather than through a constraint
\cite{rivier:1985}.
One may argue that the dynamic Voronoi random lattices also
are space-filling by construction rather than by a constraint.

Independent of temperature we find from our Monte Carlo simulation
results for the equilibrated random lattices that when plotting
$n m(n)$ against $n$  the points generally
tend to lie on slightly concave curves, as can be seen from Fig.~4(a);
however, for an
intermediate $n$ interval a linear law with a slope $\approx 5.3$
approximates the curves well.
The positive coupling
model at very low temperatures ($T=0.1$) is atypical since only points
with 5, 6 and 7 neighbours are found.
For such a random lattice with only a small number of isolated
``5''s and ``7''s one has that \mbox{$m(n) \approx 6$},
which would be exact for a triangular lattice.
This leads to \mbox{$n m(n) \approx 6 n$}, which is in accordance
with the results in Fig.~4(a) for the temperature $T = 0.1$.
Raising the temperature, points with lower and higher coordination
numbers occur and the linear law is no longer a good description.
Instead the points lie on a slightly bent curve
(cf.\  Fig.~4(a))
which can be fitted numerically by, e.g., adding a term $-bn^2$ on the
right hand side of \Eq{aboav-weaire} (cf.\ Ref.~\cite{lecaer-ho:1990}).

Also in the negative coupling case a straight line provides a good
approximation. For high $n$ values ($n_i \, \,^{>}_{\sim} \, 10$)
there are large deviations from an approximate linear law
and the $n m(n)$ values drop drastically.
Simulations on lattices with a larger number of points show
that this is due to finite-size effects, i.e., by simulating even larger
systems the curves extend to higher $n$ values before they drop.
As the temperature is lowered the curve
is shifted a little upwards but the overall form does not change
(cf.\ Fig.~4(b)).
When the temperature becomes so low that the circular patterns start
to develop we see large deviations from a linear relation with
$n m(n)$ values up to more than 100 (not shown) but, as previously discussed,
the random lattices are then difficult to fully equilibrate.

Another empirical law is Lewis' law
which has been found to hold throughout the evolution of several
two-dimensional cellular structures (see, e.g., Ref.~\cite{rivier:1985}).
Lewis' law
states that the average area $\bar{A_n}$ of $n$ sided cells
is linearily related to $n$ as
\beq
	\bar{A_n} = \bar{A} \left[ 1 + \lambda (n-6) \right]
							\label{eq:lewis}
\eeq
Here $\bar{A} = A_{tot} / N$ is the average area of a cell
belonging to a cellular network
consisting of $N$ cells (points) and with total area $A_{tot}$,
and $\lambda$ is an adjustable parameter.
The linear relation~\eq{lewis} has been obtained from maximum
entropy arguments with the constraint that the structure at all
times is space-filling \cite{rivier-lissowski:1982,rivier:1985}.
The idea underlying these maximum entropy arguments is to maximize the entropy
\mbox{$S = - \sum_n P_n \ln P_n $}, subject to the normalization constraint
\mbox{$\sum_n P_n = 1$}, the topological constraint~\eq{n=6} and the
space-filling constraint \mbox{$\sum_n P_n \bar{A_n} = \bar{A}$}.
Furthermore, it was obtained that
\mbox{$P_n = \frac{1}{3} \left( \frac{3}{4}\right)^{n-2}$}
in the $N \to \infty$ limit.
Since our probability densities (cf.\ Fig.~3) clearly are not of the above
exponential form, this means that there are other constraints
present (according to the entropy based arguments,
cf.\ Refs.~\cite{rivier-lissowski:1982,rivier:1985}).
For our system an additional constraint is
\mbox{$\sum_n P_n E_n = E(T)$}, where $E_n = (n-6)^2$ and $E(T)$
is the thermodynamic energy per point, which is measured in the
Monte Carlo simulation.

In Fig.~5 is shown the dependence of $\bar{A_n}$ on $n$ for
different temperatures for the two models (the average area
$\bar{A}$ has been normalized to unity). One notices that
Lewis' law provides a
good description of the equilibrated Voronoi random lattices and
since the random lattices can be regarded as space-filling,
this is in accordance with the results in
Refs.~\cite{rivier-lissowski:1982,rivier:1985}.
For the negative coupling model one obtains $\lambda \approx 0.23$
independent of temperature for $T \, \,^{>}_{\sim} \, 11$,
cf.\ Fig.~5(b), and then, due to the circular patterns,
large deviations (not shown) are seen where $\bar{A_n}$ for, e.g., $n=7$ and 8
can become as large as 10.  The deviations from a linear law
for high $n$ values are due to the finite number of points
and move to higher $n$ values when larger systems are simulated.
In the positive $J$ case, cf.\ Fig.~5(a), the $\lambda$ value
gradually changes from $\lambda \approx 0.23$ for high temperatures to
$\lambda \approx 0.5$ for low temperatures
with deviations from a linear law for high $n$ values (and these deviations
continue to exist for larger systems).

\section{Conclusions}

In the present paper we have investigated equilibrated random
lattices with connectivities determined by the Voronoi construction.
The random lattices have been studied using Monte Carlo simulations
where an update consisted in
moving a point to a new randomly chosen position
and then accept the new configuration with a probability depending on the
equilibrium distribution.

We have investigated the topological Hamiltonian~\eq{H=q2}
for both positive and negative couplings $J$.
We found that for negative coupling constants the low temperature phase
ressembles a glass with many metastable states, while for
positive couplings a transition
like in two-dimensional melting might be expected.

Statistical laws (Aboav-Weaire's and Lewis' laws) known to provide a
good description of natural cellular structures have been
tested for the Voronoi random lattices.
We find that the linear Aboav-Weaire law provides a good description
of equilibrated random lattices (but the points tend to lie on slightly
concave curves similar to what happens for the Poissonian random lattice),
and the linear Lewis law is followed with a temperature-dependent slope.

\section*{Acknowledgments}

K.B.L.\ gratefully acknowledges the hospitality of the
H\"ochst\-leistungs\-rechen\-zentrum (HLRZ) in J\"ulich, and the support
from the \mbox{Danish} Research Academy and the Carlsberg Foundation.

\bigskip
\noindent
\rule{5.0cm}{0.1mm}

*Permanent address:
{\em Institute of Physics and Astronomy, Aarhus University,
DK--8000 Aarhus~C, Denmark.\/} {\tt \,E-mail:\ kent@dfi.aau.dk}






%

\newpage

\section*{Figure captions}

\baselineskip=0.75cm

\begin{description}

\item[\bf FIG.\ 1:~] Part of a Voronoi random lattice (a) before and (b)
	after the point $P$ has been moved. The process appearing to
	the left of the point $P$ is a neighbour-loosing process
	whereas the process to the right is a neighbour-gaining process.
	The thick lines represent the links of the random lattice
	whereas the thin lines are the edges of the Voronoi cells.

\item[\bf FIG.\ 2:~] Random lattices and Voronoi tessellations; (a) and (b)
	obtained with the Hamiltonian~\eq{H=q2} with positive coupling $J$
	and at a temperature $T = 0.1$,
	and (c) and (d) for the corresponding negative coupling model
	after a quench to $T=5$ (and with energy $E = -$48,430).
	All the lattices contain $N=504$ points.

\item[\bf FIG.\ 3:~] Probability $P_n$ of finding a point with $n$ neighbours
	for various temperatures
	obtained with (a) the Hamiltonian~\eq{H=q2} with positive coupling $J$,
	and (b) for the corresponding negative coupling model.

\item[\bf FIG.\ 4:~] Average total number of neighbours $n m(n)$
	(Aboav-Weaire's law) for various temperatures
	obtained with (a) the Hamiltonian~\eq{H=q2} with positive coupling $J$,
	and (b) for the corresponding negative coupling model.

\item[\bf FIG.\ 5:~] Average area $\bar{A_n}$ ($\bar{A}$ normalized to unity)
	of $n$ sided cells (Lewis' law) for various temperatures
	obtained with (a) the Hamiltonian~\eq{H=q2} with positive coupling $J$,
	and (b) for the corresponding negative coupling model.

\end{description}

\end{document}